# Joint detection algorithm for multiple cognitive users in spectrum sensing

**Fanfei Meng [1, *], Yuxin Wang [1], Lele Zhang[2], Yingxin Zhao[2]**

[1]Northwestern University, Evanston, IL, United States

[2]Nankai University, 94 Weijin Rd, Nankai District, Tianjin, China

*fanfeimeng2023@u.northwestern.edu

**Abstract.** Spectrum sensing technology is a crucial aspect of modern communication technology, serving as one of the essential techniques for efficiently utilizing scarce information resources in tight frequency bands. This paper first introduces three common logical circuit decision criteria in hard decisions and analyzes their decision rigor. Building upon hard decisions, the paper further introduces a method for multi-user spectrum sensing based on soft decisions. Then the paper simulates the false alarm probability and detection probability curves corresponding to the three criteria. The simulated results of multi-user collaborative sensing indicate that the simulation process significantly reduces false alarm probability and enhances detection probability. This approach effectively detects spectrum resources unoccupied during idle periods, leveraging the concept of time-division multiplexing and rationalizing the redistribution of information resources. The entire computation process relies on the calculation principles of power spectral density in communication theory, involving threshold decision detection for noise power and the sum of noise and signal power. It provides a secondary decision detection, reflecting the perceptual decision performance of logical detection methods with relative accuracy.

## 1. Introduction

Multi-user collaborative sensing technology is developed on the basis of single-user collaborative sensing technology. The principle of multi-user collaborative sensing judgment is to use the results of each user to formulate a comprehensive judgment, without relying on the result of a single user's judgment. When the single-user perception is completed, the agent will collect the judgment results of each user and make judgments according to preset logical principles. Logical principles can be simple or complex, and the basic principles are based on maximum error tolerance and accuracy. However, due to the shortcomings of this method, it requires supervision and management of the single-user sensing process, and relies solely on the results after sensing authorization to make decisions.

    The second method is a process where the fusion of multi-user perception directly participates in the information perception and decision-making process. There are two specific approaches. One is to use the characteristics of the signal-to-noise ratio of each channel before single-user perception to pre-

estimate the allocation of signal and noise, roughly identifying potential issues that may arise during the single-user decision-making process. In the subsequent process, algorithms are utilized to minimize the impact of different channel characteristics on the single-user decision results as much as possible. Another approach is to, after single-user perception, statistically classify the results of single-user perception and propose more optimized algorithms to estimate the credibility of the entire decision-making process and integrate them. This fusion process, different from the logic gate method, involves logic gate circuits not directly or indirectly participating in the process of single-user perception but only participating in the process after the single-user perception decision.

Based on the analysis of channel spectrum resources and the discussion on how to utilize idle spectrum resources, this paper, building upon the single-user energy threshold decision method, proposes a multi-user collaborative spectrum cognitive system. This approach not only ensures the accuracy and effectiveness of the primary user analysis and decision-making but also further extends potential solutions for mitigating negative impacts from interfering users in the future.

## 2. Backgrounds and Related Work

*2.1. Spectrum sensing technology energy threshold detection model.*

The basic principle is to extract a certain width of narrowband signal, following the sampling theorem in digital signal theory. The sampling frequency is set to be greater than twice the bandwidth, ensuring distortion-free signal recovery. Subsequently, the signal intensity is squared to calculate the respective power spectra of the signal and noise. The calculated signal power spectral density needs to be added to the noise power density to serve as the decision energy basis for detection probability. Finally, the decision involves comparing the sum of noise power and false alarm threshold power with the threshold power for detection probability.

Assuming the number of samples is N, the energy E entering the detection threshold is defined as:

$$\sum_{0}^{N-1} |z(n)|^2$$

(2-1)

Suppose that the spectrum of each discrete sampled signal follows a distribution pattern of a t-distribution. Consequently, it can be expressed as the average signal-to-noise ratio $\gamma = E[|s(n)|^2]/\sigma_w^2$, where w represents the bandwidth in the time domain, and E follows a chi-square distribution. Simultaneously, the signal-to-noise ratio for each channel is constant, and if the number of samples of the spectrum is large enough, it can be assumed that the weighted power of noise and signal conforms to the characteristics of a Gaussian distribution. Therefore, the probability distribution of energy is given by:

$$E \mid H_1 \sim N(P + \sigma_n^2, \frac{2}{N}(P + \sigma_n^2)^2)$$

(2-2)

The false alarm probability can be derived as:

$$P_f = Q\left(\frac{E_{th} - \sigma_n^2}{\sqrt{2/N\sigma_n^4}}\right)$$

(2-3)

Knowing the energy detection threshold, the detection probability can also be deduced:

$$P_d = Q\left(\frac{E_{th}-(P+\sigma_n^2)}{\sqrt{2/N(P+\sigma_n^4)}}\right)$$

(2-4)

Q-function is defined as:

$$Q(x) = \frac{1}{\sqrt{2\pi}} \int_x^{+\infty} e^{-x^2/2} d\tau$$

(2-5)

Using the relationship to eliminate the threshold value, we obtain the corresponding relationship for the number of samples:

$$N = 2[Q^{-1}(P_f) - Q^{-1}(P_d)(1+SNR)]^2 SNR^{-2}$$

(2-6)

Utilizing the already defined Q function to simplify the expressions for false alarm probability and detection probability:

$$P_d = P\{E > E_{th} \mid H_1\} = Q_N\left(\sqrt{2\gamma}, \sqrt{E_{th}}\right)$$

(2-7)

$$P_f = P\{E > E_{th} \mid H_0\} = \frac{\Gamma(u \mid E_{th}/2)}{\Gamma(u)}$$

(2-8)

In the above expressions, Γ represents the incomplete gamma function, Q represents the generalized Marcum Q function, and it denotes the numerical value of the energy decision threshold used for detection. It can be observed that the noise energy in the channel determines the final detection outcome. When the noise power in the channel occupied by the detected user signal is relatively high, false alarm probability may occur.

Assuming constant noise power, and considering the scenario of Rayleigh noise characteristics in this channel, in a fading channel, the probability distribution function for the signal-to-noise ratio is assumed to follow the distribution as follows:

$$f(\gamma) = \frac{1}{\overline{\gamma}} \exp\left(-\frac{\gamma}{\overline{\gamma}}\right), \gamma \geq 0$$

(2-9)

Under this channel characteristic, the detection probability of energy can be expressed as:

$$P_d = \int_x Q_N(\sqrt{2\gamma}, \sqrt{E_{th}})$$

(2-10)

Substituting the Rayleigh channel characteristic function from the previous equation into this one, the detection probability is obtained as:

$$P_{dRay} = s^{-\frac{E_{th}}{2}} \sum_{k=0}^{N-2} \frac{1}{k!}\left(\frac{E_{th}}{2}\right) + \left(\frac{1+\overline{\gamma}}{\overline{\gamma}}\right)^{N-1} \cdot \left(e^{-\frac{E_{th}}{2(1+\overline{\gamma})}} - e^{-\frac{E_{th}}{2}} \sum_{k=0}^{N-2} \frac{1}{k!}\left(\frac{E_{th}\overline{\gamma}}{2(1+\overline{\gamma})}\right)^k\right)$$

(2-11)

*2.2. Multi-cognitive user joint wideband spectrum hard sensing technology.*
The hidden terminal problem arises due to various signal fading and interference factors, where independent channel decision users may not receive sufficient energy for judgment due to these factors [1-3]. This can lead to a decrease in decision-making perception capability. Even if the spectrum is currently occupied by a signal source, the hidden terminal problem may cause the channel decision user to erroneously believe that the spectrum is unoccupied. If, at this time, another signal source accesses the occupied channel [4-6], the original information will be severely damaged and affected, making recovery difficult. The hidden terminal problem is one of the significant challenges that spectrum sensing technology in mobile communication needs to address.

In the depicted scenario, a centralized mobile communication sensing tree network covers an information synthesis fusion and decision center and three independently channel-aware decision users CR1, CR2, and CR3[7]. The information transmission channel between the primary user and CR1 is obstructed by a large building, causing severe large-scale fading and signal attenuation. If only one independent channel-aware decision user is used for CR1[8-9], it is very likely that the presence of the primary signal cannot be perceived, giving rise to the hidden terminal problem [10-12]. If CR1 accesses the channel while the primary user is active, it will cause interference to the primary user [13].

### 3. Algorithm and Methodology

*3.1. "AND" logistic principle.*
Under the "AND" logic criterion, the fusion center releases a comprehensive result after synthesizing and analyzing the perception results of all secondary users. In other words, all independently channel-aware decision users must have consistent results for the conclusion to be valid. If there are N independent channel decision-makers, the final result after fusion perception analysis based on the "AND" logic criterion is given by:

$$Q_d = \prod_{i=1}^{N} P_{d,i}$$

(2-12)

$$Q_f = \prod_{i=1}^{N} P_{f,i}$$

(2-13)

Where $P_{d,i}$ represents detection probability, $P_{f,i}$ represents False alarm probability.

*3.2. "OR" logistic principle.*
Each independent perceptual decision-maker conducts a comprehensive analysis through the logical "OR." As long as one cognitive user detects that the channel is occupied, the fusion perception center considers the spectrum to be occupied. Assuming there are a total of N cognitive users, it can be deduced that this logical criterion can yield the overall false alarm probability and detection probability:

$$Q_d = 1 - \prod_{i=1}^{N} (1 - P_{d,i})$$

(2-14)

$$Q_f = 1 - \prod_{i=1}^{N} (1 - P_{f,i})$$

(2-15)

It can be seen that this logical relationship decision criterion imposes very stringent requirements for determining the spectrum unoccupied. It is necessary for all cognitive users to believe that the spectrum is unoccupied before concluding that the spectrum is idle.

*3.3. K-rank principle.*
The "K-rank" criterion is based on the belief that among N independent channel-aware decision users,

at least K users must provide feedback that the spectrum is occupied; otherwise, the channel is considered unoccupied. It can be observed that when K equals N, the "K-rank" criterion is similar to the AND logical condition, and when K equals 1, the "K-rank" criterion is similar to the OR logical detection condition. Therefore, it can be summarized with the following relationship:

$$Q_d = \sum_{i=K}^{N} C_N^i \prod_{j=1}^{i} P_{d,j} \prod_{m=1}^{} (1 - P_{d,m})$$

(2-16)

$$Q_f = \sum_{i=K}^{N} C_N^i \prod_{j=1}^{i} P_{d,j} \prod_{m=1}^{} (1 - P_{f,m})$$

(2-17)

Generally, when greater emphasis is placed on the accuracy of the false-negative probability, we raise the detection threshold value in the OR criterion case to optimize detection performance. Conversely, in the AND criterion system, we lower the detection threshold value to achieve better detection performance in situations where the emphasis is on the opposite.

*3.4 Soft decision data fusion and its security risks*

Each independent channel decision user sends decision information to the fusion decision center in the form of 1-bit information {0, 1}, where "1" and "0" respectively indicate whether the channel of the signal source is occupied (H1) or not occupied (H0). This data format keeps the channel occupancy low during the decision transmission process, but the detection performance is not very ideal.

In comparison with the method in 2.1.1, the fusion center does not wait until independent channel decision users have completed their cognitive processes before conducting unified comprehensive analysis. Instead, it joins the observation during the decision-making process of independent cognitive users, using collaborative diversity methods to perceive the collected results. Therefore, soft decision algorithms can have superior cognitive capabilities and accuracy. Among many soft decision perception algorithms, the Sequential Probability Ratio Test (SPRT) is the most common, with a broad range of applications. Its basic principle involves analyzing the randomly sampled prior probabilities P(di|H1) and P(di|H0) of the data information di collected by each independent channel-aware decision user under the events H1 and H0, and calculating the decision statistic Sn.

$$S_n = \prod_{n}^{i=0} \frac{P(d_i \mid H_1)}{P(d_i \mid H_0)}, n = 1, 2, 3...$$

(2-18)

The advantage of the Sequential Probability Ratio Test (SPRT) is that, while ensuring the same spectrum detection performance, the number of detection samples is not fixed, making the time cost lower compared to requiring a fixed number of detection samples. However, the good performance of the SPRT algorithm is contingent on the assumption that cognitive detection decision users are friendly, lacking a certain degree of self-protective security mechanism. The vulnerability in self-protection and susceptibility to malicious interference to some extent make the SPRT algorithm prone to SSDF malicious interference from malicious users, suppressing its judgment capability for the available spectrum status of the main user.

Based on the target of malicious attacks, there are two types of SSDF malicious interference: Always-free and Always-false. The first type is Always-busy malicious interference: the interferer continuously sends false data indicating the occupied status of the main user's spectrum. The fusion center may mistakenly believe that the main user is always in a "busy" state, allowing malicious attackers to illegitimately occupy idle spectrum resources for an extended period. The second type is Always-free attack: malicious users continuously send false data indicating the unoccupied status of the main user's spectrum. The fusion center may mistakenly believe that the main user is always in an "idle" state, allowing illegal malicious users to interfere with the main user's transmitted signal. Currently, another algorithm, the TNA algorithm, utilizes a trust node-assisted secure cooperative perception scheme to eliminate the negative effects of interferers on the SPRT soft decision algorithm. However, its effectiveness against the interferer's sporadic intermittent SSDF interference actions is not obvious.

According to the interferer's interference mode, there are two types of SSDF interference: persistent and intermittent. In the case of persistent SSDF interference actions, the interferer continuously uploads false perception information, resulting in very low credibility and making it extremely easy for the fusion center to identify. Consequently, the interferer switches interference modes, intermittently uploading decision information to dynamically fabricate false behavior. In the alternating upload process of real or false perception information, the interferer strategically keeps it in a credible state, thereby avoiding detection and deepening the threat to cooperative spectrum perception decisions.

In response to the security risks of soft decision data fusion, some scholars have proposed a trust mechanism based on feedback reputation. First, establish a network model for feedback, carefully construct the communication of cognitive information and decision information among multiple independent channel users in cooperative cognitive comprehensive decisions. Second, add individual-affirmation counts and individual-complaint counts to evaluate the trust value of individual feedback decision data. Finally, using the trust value evaluation system for feedback decision data, the quantification process of reputation values is completed.

## 4. Experiments

*4.1. Single-user energy threshold spectrum perception.*
This section first employs the method of single-user energy threshold to make judgments about the spectrum situation within the main user, followed by an analysis of the improvement and optimization in the collaborative perception process of multiple users. In this simulation, three channels were set up, each with a signal-to-noise ratio of 5 dB, -8 dB, -10 dB. The three channels are mutually independent without interference and make relatively independent decisions. In the first stage, each single user performs spectrum perception separately, and the false alarm probability curve and detection probability curve for each decision are plotted.

In the simulation, assuming the signal bandwidth in this frequency band is $W=5*10^4$, according to Nyquist sampling theorem and Monte Carlo theorem, the sampling frequency $F=10^5$ Hz. Therefore, the period of the sampled signal is T=0.001. Throughout the process, the number of repetitions is set to 5000 times, and the noise and signal power of the channel are fixed. The energy thresholds for the three channels are set as follows: 200600, 500900, 700~1300, increasing by 20 at each step.

Calculation method for false alarm probability: Compare the noise power with the energy value of the current threshold during each repetition. If the noise power is greater than the current threshold power, then add 1 to the false alarm probability statistical value. The ratio of the false alarm probability detection value to the total number of repetitions is the false alarm probability.

Calculation method for detection probability: First, compare the noise power with the energy value of the current threshold during each repetition. If the noise power is less than the current threshold power, then compare whether the sum of the signal power and noise power is greater than the energy threshold. If it is, then add 1 to the detection probability statistical value. The ratio of the detection probability detection value to the total number of repetitions is the detection probability.

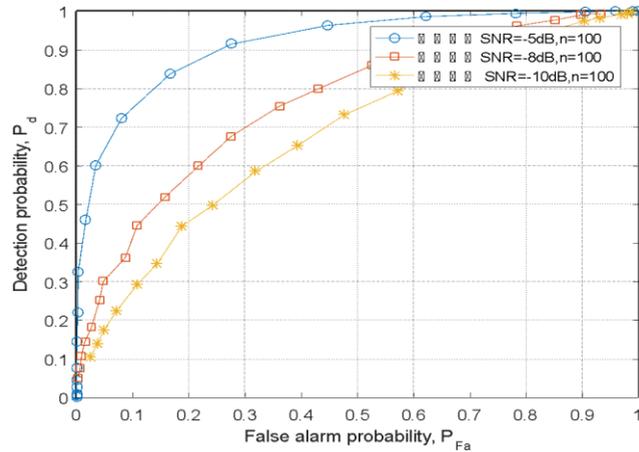

**Figure 1**: Schematic diagram of single-user multiple-channel independent detection decisions

As shown in Figure 1, it illustrates the independent decision results of single-user energy detection thresholds for each channel. This decision result is primarily based on the curve relationship between false alarm probability and detection probability. The higher the detection probability, the more accurate the decision result, and the steeper the slope of the curve. Therefore, from the graph, it can be observed that when the signal-to-noise ratio of the channel is relatively low, the method of single-user energy detection threshold has a good decision effect. In this situation, the power of the noise is relatively small, resulting in lower interference from the noise to the main signal. This makes the method of energy detection decision threshold have excellent cognitive capabilities.

*4.2. Multiple-User Energy Threshold Spectrum Perception.*

In this section, the spectrum situation within the main user is still judged using the energy threshold method, but it extends to multi-user collaborative spectrum perception decisions based on multiple single users. The simulation still involves three channels, each with signal-to-noise ratios of 5 dB, -8 dB, -10 dB. The three channels are mutually independent without interference and make relatively independent decisions. In the first stage, each single user performs spectrum perception separately. Then, the fusion center collects the perception results of the three channels and makes a fused perception decision to determine the final cognition of the occupation status of the main user's channel. Before the fusion center obtains the perception results, the three independent channel-aware users do not exchange information with each other. The results of independent perception are unique to each channel and are not influenced by the decision results of other channels.

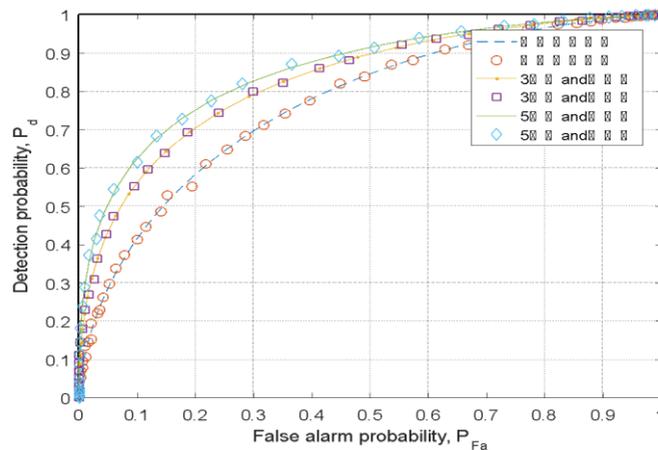

**Figure 2**: Schematic diagram of multiple-channel independent detection decisions under "AND" logistics

As shown in Figure w, the decision results of independent channels with energy detection thresholds for both five users and three users are presented, with a signal-to-noise ratio of -8 dB for each of them.

This decision method requires that all users participating in the decision must believe that there is a signal in the channel for the fused decision to conclude that the main user's channel spectrum is occupied. The solid line and dashed line represent the theoretical and simulated results, and the difference between the two is not very large. This indicates that the simulation fitting is relatively ideal. The decision results are primarily based on the relationship between false alarm probability and detection probability. The higher the detection probability, the more accurate the decision result, and the steeper the slope of the curve.

From the graph, it can also be observed that when the signal-to-noise ratio of the channel is relatively low, the decision effect of the multi-user energy detection threshold method is still good. In this situation, the power of the noise is relatively small, resulting in lower interference from the noise to the main signal. This makes the method of energy detection decision threshold have excellent cognitive capabilities. Compared to the spectrum perception method with a single-user energy threshold, the contrast between the two results clearly shows that the method of multi-user collaborative spectrum perception decision has a higher slope of the curve for detection probability relative to false alarm probability, indicating a better decision effect.

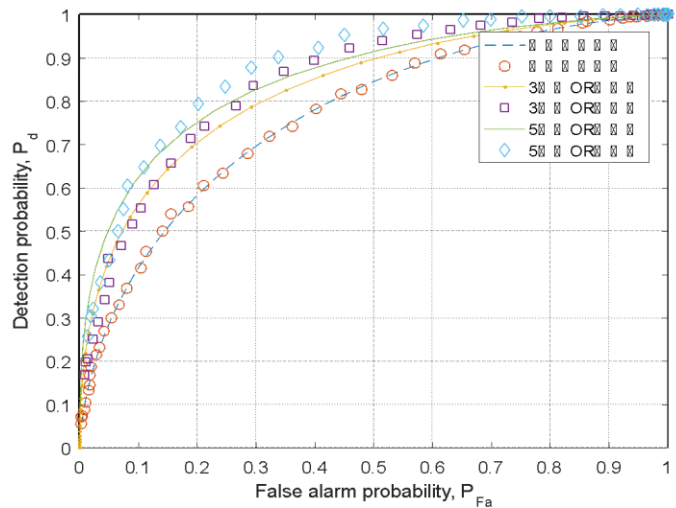

**Figure 3**: Schematic diagram of multiple-channel independent detection decisions under "OR" logistics

As shown in Figure 3, the decision results of independent channels with energy detection thresholds for both five users and three users are still presented, with a signal-to-noise ratio of -8 dB for each of them. This decision method requires that as long as one user participating in the decision believes that there is a signal in the channel, the decision can be made that the main user's channel spectrum is occupied. In terms of decision accuracy, the slope changes of the false alarm probability and detection probability curves are generally similar. However, the "OR" logic criterion has a higher detection probability. This indicates that when the decision requirements are relaxed, there will be a corresponding performance improvement in detection probability.

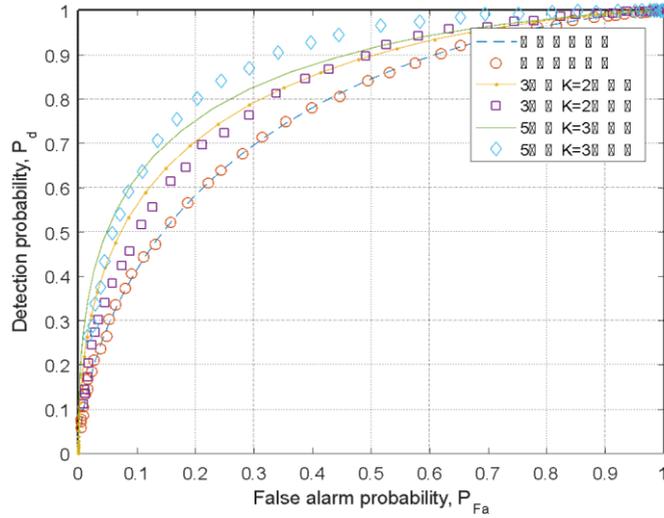

**Figure 4**: Schematic diagram of multiple-channel independent detection decisions under "K-rank" logistics

As shown in Figure 4, the decision results of independent channels with energy detection thresholds for both five users and three users are still presented, with a signal-to-noise ratio of -8 dB for all channels. According to the "K-rank" criterion mentioned earlier, when making decisions with three users, setting K=2 means that if two or more users participating in the decision believe that there is a signal in the channel, the decision can be made that the main user's channel spectrum is occupied. When making decisions with five users, setting K=3 means that if three or more users participating in the decision believe that there is a signal in the channel, the decision can be made that the main user's channel spectrum is occupied. In terms of decision accuracy, the slope changes of the false alarm probability and detection probability curves are still generally similar to the previous two simulations. In a more detailed analysis, its accuracy performance is between the "AND" logic criterion and the "OR" logic criterion. This indicates that when the decision criteria are between the two, the detection performance is also between the two.

## 5.    Conclusions

While the method of using energy thresholds to achieve spectrum awareness has been essentially realized and three corresponding perception results for different decision logics have been studied, there are still possibilities for improvement and refinement in the simulation of this paper:

*1. Nonlinear Channel Considerations:* In scenarios where the channel exhibits nonlinear characteristics, the actual signal-to-noise ratio (SNR) cannot be fixed in real-time. This leads to a significant fluctuation range in the estimation of noise values.

*2. Limitations of Gaussian Channel Models:* Gaussian channel models have limitations, particularly in real-world applications where they may not accurately represent actual channel characteristics, especially when large-scale fading occurs. Exploring more sophisticated models that capture the complexities of real-world channel features could improve the accuracy of the simulation.

*3. Challenges of Single Floating Energy Threshold Setting:* Setting a single floating energy threshold is inherently challenging and can result in higher false alarm probabilities and lower detection probabilities. Dual-threshold energy detection not only helps address issues where signal power and noise power are close but also provides good mathematical differentiability and extendability for subsequent mathematical analyses in fusion perception.

*4. Error Considerations in Logic Circuit Design:* The design of multi-user decision-making methods involving logic circuits may encounter issues related to circuit errors. In this context, errors in decision-making may not necessarily originate from individual decision-makers but can be attributed to

DOI: 10.54254/2977-3903/4/2023053

errors in the logic circuit itself. This can introduce unpredictable errors in drawing dashed lines for false alarm probabilities and detection probabilities.

Addressing these considerations could contribute to a more accurate and robust simulation of spectrum awareness in the presented framework.